\documentclass[bimj,fleqn]{w-art}
\usepackage{times}
\usepackage{w-thm}
\usepackage[colorlinks,urlcolor=black,citecolor=black,linkcolor=black]{hyperref} % ,pagebackref
\usepackage{orcidlink}
\usepackage[authoryear]{natbib}
\usepackage{amsmath,amssymb} % amsfonts
\usepackage{amsthm,amscd} % Ths usw
\usepackage{graphicx,color}
\usepackage{url}
\usepackage{booktabs} % [c]midrule
\usepackage{setspace}

\theoremstyle{plain}

\theoremstyle{definition}

\usepackage[]{graphicx}
\chardef\bslash=`\\ % p. 424, TeXbook

\DeclareMathOperator{\dCov}{dCov}

\newcommand{\dCovh}{\widehat{\dCov}}

\DeclareMathOperator{\E}{E}
\DeclareMathOperator{\Prob}{P}

\DeclareMathOperator{\Cov}{Cov}

\DeclareMathOperator{\Normal}{\mathcal{N}}
\DeclareMathOperator{\Multin}{Multinomial}
\DeclareMathOperator{\Multib}{Multi-Bernoulli}

\DeclareMathOperator{\tr}{tr}
\DeclareMathOperator{\rank}{rank}
\DeclareMathOperator{\vecop}{vec}
\newcommand{\arr}{\longrightarrow}

\newcommand{\ceil}[1]{\left\lceil{#1}\right\rceil}
\newcommand{\brc}[1]{\left\lbrace{#1}\right\rbrace}
\newcommand{\detlim}{\underset{n\to\infty}{\arr}}

\newcommand{\aslim}{\overset{a.s.}{\detlim}}
\newcommand{\distrilim}{\stackrel{\mathcal{D}}{\longrightarrow}}

\newcommand{\tensor}{\otimes}

\newcommand{\transp}{^\text{t}}

\newcommand{\R}{\mathbb{R}}
\newcommand{\Rplus}{\mathbb{R}^{+}}

\newcommand{\Zplus}{\mathbb{Z}^{+}}
\newcommand{\ba}{\mathbf{a}}
\newcommand{\bA}{\mathbf{A}}
\newcommand{\bB}{\mathbf{B}}
\newcommand{\bc}{\mathbf{c}}
\newcommand{\bC}{\mathbf{C}}
\newcommand{\bD}{\mathbf{D}}
\newcommand{\bE}{\mathbf{E}}
\newcommand{\bG}{\mathbf{G}}
\newcommand{\bH}{\mathbf{H}}
\newcommand{\bI}{\mathbf{I}}
\newcommand{\bM}{\mathbf{M}}
\newcommand{\bp}{\mathbf{p}}
\newcommand{\bq}{\mathbf{q}}
\newcommand{\bQ}{\mathbf{Q}}
\newcommand{\br}{\mathbf{r}}
\newcommand{\bU}{\mathbf{U}}
\newcommand{\bV}{\mathbf{V}}
\newcommand{\bw}{\mathbf{w}}

\newcommand{\bGamma}{\mathbf{\Gamma}}
\newcommand{\bLambda}{\mathbf{\Lambda}}
\newcommand{\bzero}{\mathbf{0}}
\newcommand{\bone}{\mathbf{1}}
\newcommand{\eps}{\varepsilon}

\newcommand{\spx}{\mathcal{X}}
\newcommand{\spy}{\mathcal{Y}}
\newcommand{\spz}{\mathcal{Z}}
\newcommand{\dx}{d_{\spx}}
\newcommand{\dy}{d_{\spy}}

\begin{document}
\DOIsuffix{bimj.70129}
\Volume{68}
\Issue{3}
\Year{2026}
%\Page{e70129}
\pagespan{1}{20}

\title[Distance covariance for categorical data]{Tests for categorical data beyond Pearson: A distance covariance and energy distance approach}

%%%% AUTHORS %%%%
\author[Castro-Prado {\it{et al.}}]{Fernando Castro-Prado\orcidlink{0000-0003-3722-2013}~\footnote{Corresponding author: {\sf{e-mail: fernando.castro.prado@alumni.dkfz.de}}
}\inst{,1,2,3}}

\author[]{Wenceslao Gonz\'alez-Manteiga\orcidlink{0000-0002-3555-4623}\inst{1}}

\author[]{Javier Costas\orcidlink{0000-0003-0306-3990}\inst{2}}

\author[]{\\Fernando Facal\orcidlink{0000-0002-1157-6772}\inst{2}}

\author[]{Dominic Edelmann\orcidlink{0000-0001-7467-6343}\inst{3}}

%%%% POSTAL ADDRESSES %%%%
\address[\inst{1}]{Department of Statistics, Faculty of Mathematics, University of Santiago de Compostela, R\'ua Lope G\'omez de Marzoa s/n, 15782 Santiago de Compostela, Galicia, Spain.}

\address[\inst{2}]{Psychiatric Genetics Laboratory, Santiago Health Research Institute (IDIS), University Hospital, Travesía da Choupana s/n, 15706 Santiago de Compostela, Galicia, Spain.}

\address[\inst{3}]{Biostatistics Department, German Cancer Research Center (DKFZ), Im Neuenheimer Feld 280, 69120 Heidelberg, Baden-Württemberg, Germany.}

\Receiveddate{15 January 2024} \Reviseddate{20 July 2024} \Accepteddate{16 October 2024}

\begin{abstract}
% One paragraph. Max.: 250 words. This is ~170 words.

Categorical variables are of uttermost importance in biomedical research. When two of them are considered, it is often the case that one wants to test whether or not they are statistically dependent. We show weaknesses of classical methods ---such as Pearson's and the $G$-test--- and we propose testing strategies based on distances that lack those drawbacks. We first develop this theory for classical two-dimensional contingency tables, within the context of distance covariance, an association measure that characterizes general statistical independence of two variables. We then apply the same fundamental ideas to one-dimensional tables, namely to the testing for goodness of fit to a discrete distribution, for which we resort to an analogous statistic called energy distance. We prove that our methodology has desirable theoretical properties, and we show that we can calibrate the null distribution of our test statistics without resampling. We illustrate all this in simulations, as well as with some real data examples, demonstrating the adequate performance of our approach for biostatistical practice.

\end{abstract}

\keywords{Distance covariance; contingency tables; independence testing; categorical data; Pearson's chi-squared test.\\[1pc]}

\maketitle

\section{Introduction}

In previous work by us \citep{Epistasis:paper}, an interesting data set from complex disease genomics motivated us to define distances on discrete spaces of cardinality 3 and test independence among variables whose support lies on such spaces. Since the times of Karl Pearson (more than a century ago), the corresponding test for categorical variables with an arbitrary finite number of categories has been of paramount interest to manifold applications. As a matter of fact, independence of categorical variables ranks among the most often tested hypotheses in biomedical practice \citep{BS}. Discrete data arise in health sciences in a variety of contexts \citep{Agresti,Preisser} --- for measuring responses to treatments, signposting the stage of a disease (or whether the disease is present), establishing subgroups after a diagnosis, and so forth.

In this paper, we present the distance and kernel counterpart \citep{DJ} of what \citet{Pearson} did. We derive some theory for independence testing and extend it to the problem of goodness of fit. We finally illustrate the performance of our methodology with synthetic and real data examples, including the comparison with competing methods.

For independence, we will consider categorical variables $X\in\{1,\ldots,I\}$ and $Y\in\{1,\ldots,J\}$. Given an IID sample $\{(X_m,Y_m)\}_{m=1}^n$, one can construct the $I\times J$ contingency table $(n_{ij})_{i,j}$ by counting the observations per pair of categories $(X,Y)$:
$$
n_{ij} = \sum_{m=1}^n 1_{\{X_m = i, Y_m = j \}}.
$$
Under the null hypothesis, we expect to observe, in each cell:
$$
n_{ij} ^*:=\frac{1}{n} \sum_{k=1}^J n_{ik} \sum_{k=1}^I n_{kj}\;\:.
$$
One of the most common test statistics is Pearson's:
$$
\chi^2=\sum_{i=1}^{I} \sum_{j=1}^{J}  \frac{(n_{ij} - n^*_{ij})^2}{n^*_{ij}},
$$
for which the $p$-values are either computed using a chi-squared distribution with $(I-1)(J-1)$ degrees of freedom, or with permutations. The same holds for the null distribution of the $G$-test:
$$
G=2\sum_{i=1}^{I} \sum_{j=1}^{J}  n_{ij}\log\left( \frac{n_{ij}}{n^*_{ij}}\right),
$$
which is essentially the likelihood ratio test for this problem \citep[][Section~2.4.1]{Agresti}. Other available methods include Fisher's exact test \citep{FET} and the $U$-statistic permutation test \citep{BS}. The authors of this last work very illustratively show how classical methods have important limitations related to imbalanced cell counts, which justifies the need for new techniques for such a relevant problem.

For the problem of goodness of fit, it is customary to resort to Pearson's (chi-squared) test, for which the philosophy is, once more, ``the squared difference of the observed and the expected, divided by the expected;'' now with the difference that the table is $1\times I$ and the expected cell counts will be: $$n_i^*=n\Prob_{H_0}\{X=i\}\;.$$

The scope of this work will be to address the testing for independence and goodness of fit with categorical data, using the aforementioned techniques, collectively known as \emph{energy statistics} \citep{TEOD}. The remainder of the article is organized as follows. Section 2 contains our novel approach to the testing for independence between two categorical variables. In Section 3, we develop the testing for goodness of fit to a discrete distribution using the same basic notions, but with different theoretical tools. Some illustrative simulations are reported in Section 4. In Section 5, we apply the method to real data, to show applicability. Concluding remarks are given in Section 6. Proofs for our theoretical results are given in appendices A and B.

\section{The distance covariance test of independence between two categorical variables}\label{test:dc:categorical}

Given an IID sample $\{(X_m,Y_m)\}_{m=1}^n$ of $(X,Y)$, a consistent (but biased) estimator for the generalized distance covariance \citep{TEOD} between our two jointly distributed random variables is given by
$$
\widehat{V} = \widehat{T}_1 - 2 \widehat{T}_2 + \widehat{T}_3,
$$
where
\begin{align*}
	\widehat{T}_1 &= \frac{1}{n^2} \sum_{l,m =1}^n \dx(X_l,X_m) \, \dy(Y_l,Y_m),\\
	\widehat{T}_2 &= \frac{1}{n^3} \sum_{l =1}^n  \big( \sum_{m =1}^n \dx(X_l,X_m) \big) \, \big( \sum_{m =1}^n \dy(Y_l,Y_m) \big) , \\
	\widehat{T}_3 &= \frac{1}{n^4} \Big(\sum_{l,m =1}^n \dx(X_l,X_m) \Big) \, \Big(\sum_{l,m =1}^n \dy(Y_l,Y_m) \Big).
\end{align*}

We assume that the supports $\spx$ and $\spy$ of $X$ and $Y$ respectively are finite, with cardinality $I\in\Zplus$ and $J\in\Zplus$. When it comes to deciding which (pseudo)metrics $\dx$ and $\dy$ to equip them with, the only restriction we have for distance covariance and associated techniques to work out is that we need to be in a (pseudo)metric structure of strong negative type \citep{FW}. Now the question would be which of those feasible distances is the most convenient to use. Since we are working with categorical data and we want to be as agnostic as possible in terms of the underlying relationships among categories, in the following we will restrict ourselves to the case in which the metric structure on both marginal spaces reflects this agnosticism. In other words, we will equip both $\spx$ and $\spy$ with the discrete distance (which we will henceforward denote simply as $d$ for both spaces):
$$
d(z,z') = 1- \delta_{z \, z'}=1_{ \{z\neq z'\} }
$$
where $\delta_{\cdot \, \cdot}$ denotes the Kronecker delta and $(z,z')$ is either in $\spx\times\spx$ or in $\spy\times\spy$. Alternatively, we could obtain the same test statistic by identifying the $I$ categories of $X$ with an orthonormal basis of $\mathbb{R}^I$ and then using the Euclidean distance and classical distance covariance \citep{SRB}, instead of its extension to metric spaces \citep{Jakobsen,Lyons}.

We now construct the $I \times J$ contingency table for the sample $\{(X_m,Y_m)\}_{m=1}^n$ of $(X,Y)$. Its $(i,j)$-th cell will be denoted by $n_{ij}$:
$$
n_{ij} = \sum_{m=1}^n 1_{\{X_m = i, Y_m = j \}}.
$$
We call the $n_{ij}$'s \emph{observed} cell counts, whereas their \emph{expected} counterparts are their expected values under the null hypothesis (i.e., independence of $X,Y$).

We now introduce the notation $n_{i \cdot}$ and $n_{ \cdot j}$ for the row and column sums of the contingency table:
$$
n_{i \cdot} := \sum_{j=1}^J n_{ij} = \sum_{m=1}^n 1_{\{X_m = i\}};
$$
$$
n_{\cdot j} := \sum_{i=1}^I n_{ij} = \sum_{m=1}^n 1_{\{Y_m = j\}}.
$$	
These allow us to define the expected cell counts (under independence):
$$
n^*_{i j} = \frac{1}{n} n_{i \cdot} n_{\cdot j}
$$
By performing some algebraic manipulations, one can see that our test statistic can compactly be written as:
\begin{equation}\label{ts:indep}
\widehat{V} =\frac{1}{n^2}  \sum_{i=1}^{I} \sum_{j=1}^{J}  (n_{ij} - n^*_{ij})^2
\end{equation}

On the other hand, Pearson's (chi-squared) test for independence is based on the statistic 
$$
\chi^2=\sum_{i=1}^{I} \sum_{j=1}^{J}  \frac{(n_{ij} - n^*_{ij})^2}{n^*_{ij}},
$$
which only differs in a ``normalizing'' denominator in each term of the sum.

We now state the following result on the null distribution of our test statistic~(\ref{ts:indep}). A self-contained proof can be found in Appendix A, but we would like to point out that general theory for degenerate $V$-statistics \citep{de Wet} also allows to derive an asymptotic null distribution for the same statistic. 

\begin{theorem} \label{th:indep}
	Let $(X_1,\ldots,X_n)$ and $(Y_1,\ldots,Y_n)$ be IID samples of jointly distributed random variables $(X,Y) \in \{1,2,\ldots,I\} \times\{1,2,\ldots,J\}$, with $q_i := P(X=i)$ and $r_j := P(Y=j)$.
	
	Consider $\mathcal X$ and $\mathcal{Y}$ equipped with the discrete metric. Then the empirical distance covariance between the two random variables can be written as:
    $$
        \dCovh_{\text{discrete}}^2(X,Y) = \frac{1}{n^2}  \sum_{i=1}^{I} \sum_{j=1}^{J}  (n_{ij} - n^*_{ij})^2
    $$
	
	In addition, whenever $X$ and $Y$ are independent, for $n \to \infty$,
	$$%\begin{equation}\label{eq:asym:dc}
	n \, \dCovh_{\text{discrete}}^2(X,Y)
	\distrilim 
	\sum_{i=1}^{I-1} \sum_{j=1}^{J-1} \lambda_i \mu_j Z_{ij}^2
	$$%\end{equation}
	
	where $Z_{ij}^2$ are independent chi-squared variables with one degree of freedom each. $\lambda_1,\ldots,\lambda_I$ are the eigenvalues of matrix $\mathbf A=(a_{ij})_{I\times I}$, whose entries are:
	$$
        a_{ij} = q_i \delta_{ij} - q_i q_j,
	$$
 where $\delta_{ij}$ is the Kronecker delta.
	Similarly, $\{\mu_1,\ldots, \mu_J\}$ is the spectrum of  $\mathbf B=(b_{ij})_{J\times J}$, with
	$$
        b_{ij} = r_i \delta_{ij} - r_i r_j.
	$$
\end{theorem}

It should be noted that $\mathbf A$ and $\bB$ are the covariance matrices of a multinomial distribution multiplied by a factor (actually, of a ``multi-Bernoulli'' distribution).% \textcolor{blue}{Add the references I found on Cov matrices of Multinomials}

In practice, when it comes to using the distribution above, we will take the empirical estimators $\hat q_i$ and $\hat r_j$, then construct estimators of $\bA$ and $\bB$ from them, to finally use the products of their eigenvalues as the coefficients in the linear combination of IID $\chi^2_1$'s.

Hence, obtaining the $p$-values of our test boils down to evaluating the distribution function of weighted sums of chi-squared variables. The approximation of quadratic forms of Gaussian variables has been very well studied historically and it arises fairly often in statistical practice \citep{Duchesne}. The algorithm by \citet{Imhof} is arguably one of the best known ones, but its speed can come at the price of precision \citep{Jelle:Biomet}. We have instead chosen to resort to \citet{Farebrother} for our approximations, in the implementation by \citet{Duchesne}.

\section{The energy test for goodness of fit to a discrete distribution}\label{test:ed:categorical}
Let us once again consider a categorical variable $X$ with support $\spx$ of cardinality $I\in\mathbb Z^+$, which we will assume to be $\{1,\ldots,I\}$ without loss of generality. We observe a sample $X_1,\ldots,X_n$ IID $X$ and we will use it to test for $X\sim F$ having been drawn from a certain distribution $F_0$:
$$
H_0:F= F_0
$$

The distance-based statistic for this kind of test would be the adaptation of the one by \citet{N} to our setting. Let $d$ denote once more the discrete distance on the support of $X$. Then, the energy distance between the empirical distribution and $F$ (which equals $F_0$ under the null hypothesis) is:
$$
\mathcal E_n=n\left[ \frac{2}{n}\sum_{l=1}^n\E d(x_l,X)-\E d(X,X') -\frac{1}{n^2}\sum_{l,m=1}^n d(x_l,x_m) \right] ;
$$
where $\{x_l\}_{l=1}^n$ is a sample realization of $\{X_l\}_{l=1}^n$ and $X'$ is an IID copy of $X$. We refer the reader to \citet{ED} for a more comprehensive review on this kind of statistics.

If we now define $p_i:=\Prob_{H_0}\{X=i\}$ (for $i=1,\ldots,I$), we have that the expected cell count for each category is $n_i^*:=n p_i$, whereas the observed cell count is simply:
$$
n_i:=\sum_{l=1}^n 1_{ \{X_l=i\} }
.$$
With this notation, and after some algebra, we can write our test statistic for $H_0:F= F_0$ as:
$$
\mathcal E_n=\frac{1}{n}\sum_{i=1}^I(n_i-n_i^*)^2,
$$
which again resembles Pearson's without its denominator. As of its null distribution, we present the following result.

\begin{theorem} \label{th:gof}
	Let $(X_1,\ldots,X_n)$ be an IID sample of random variable $X \in \mathcal X=\{1,2,\ldots,I\}$.

Consider $\mathcal X$ equipped with the discrete metric. Then the energy distance test statistic for goodness of fit to a fixed distribution $\bp=(p_i)_{i=1}^I$ on $\{1,\ldots,I\}$ is:
$$
\mathcal E_n=\frac{1}{n}\sum_{i=1}^I(n_i-n_i^*)^2,
$$
with the observed counts being $n_i:=\sum_{l=1}^n 1_ { \{X_l=i\} } $ and the expected ones: $n_i^*=np_i$.

Then, whenever $X$ is distributed according to $\bp$, for $n \to \infty$,
$$
\mathcal E_n
\distrilim 
\sum_{i=1}^{I-1} \lambda_i Z_{i}^2
$$
where $Z_{i}^2$ are independent chi-squared variables with one degree of freedom each. $\lambda_1,\ldots,\lambda_I$ are the eigenvalues of matrix $\bC=(c_{ij})_{I\times I}$ with
$$
c_{ij} = p_i \delta_{ij} - p_i p_j,
$$
where $\delta_{ij}$ is the Kronecker delta.
\end{theorem}

Note that, matrix $\bC$ here is, once again, a covariance matrix of a multinomial, and therefore has zero as one of its eigenvalues and $I-1$ as its rank.

For the proof of the preceding theorem, we forward the reader to Appendix B.

\section{Simulation study}\label{simu:categorical}

We will now show how the tests proposed in Sections~\ref{test:dc:categorical} and~\ref{test:ed:categorical} perform numerically, by simulating some population models that we consider illustrative. Subsection~\ref{simu:dc} is devoted to the distance-covariance test and Subsection~\ref{simu:ed}, to the one based on the energy distance.

\subsection{Distance-covariance test of independence}\label{simu:dc}

As previously mentioned, the test statistic we present in Section~\ref{test:dc:categorical} is (almost) the same as the USP test statistic by \citet{BS}, with the substantial ---albeit not fundamental--- difference being that theirs is the $U$-statistic counterpart of our $V$-statistic. The approach for the testing, however, is completely different, since they use permutations, whereas we derive the (asymptotic) null distribution of the test statistic (Theorem~\ref{th:indep}). Nevertheless, applying classical $U$-statistic theory \citep[][Section 3.2.2]{Lee}, one can see that the USP test statistic (times the sample size) has as its asymptotic null distribution the same quadratic form described in Theorem~\ref{th:indep} for distance covariance.%Equation~\eqref{eq:asym:dc}.

We will therefore use the family of models for contingency tables with exponentially decaying marginals described by \citet{BS}, as it provides a good framework for assessing both the calibration of significance and the performance in terms of power. We will compare our method with theirs, as well as with the asymptotic distribution of the USP, Pearson's chi-squared test, Pearson's test with permutations, Fisher's exact test and the $G$-test.

Let us first define the model. For fixed $I$ and $J$, we define the cell probabilities of our contingency table under independence as:
$$
p_{ij}^{(0)}:=\frac{2^{-(i+j)}}{(1-2^{-I})(1-2^{-J})}\text{; for }i=1,\ldots,I; j=1,\ldots, J.
$$
The above expression is clearly the product of the marginal probabilities. It is also easy to see that the probability mass is maximized in the top-left corner of the contingency table and it decreases rightwards and downwards.

Now, for each $\eps\in\Rplus$ small enough so that no probabilities are out of $[0,1]$, we define $p_{ij}^{(\eps)}$ as the following perturbation of $p_{ij}^{(0)}$:
$$
p_{ij}^{(\eps)}:=\begin{cases}
	p_{ij}^{(0)}+\eps&\text{if }(i,j)\in\{(1,1),(2,2)\}\\
	p_{ij}^{(0)}-\eps&\text{if }(i,j)\in\{(1,2),(2,1)\}\\
	p_{ij}^{(0)}&\text{otherwise }
\end{cases};
$$
where $\eps\leq\min\brc{ \left[ 8(1-2^{-I})(1-2^{-J}) \right]^{-1}, 1-\left[ 4(1-2^{-I})(1-2^{-J}) \right]^{-1} }$. The larger $\eps$ is (within its range), the further the contingency table is from the null hypothesis. The upper bound for $\eps$ can be arbitrarily close to $0.125$ (as both $I$ and $J$ tend to infinity), but for us it will be approximately $\frac{1024}{7905}\approx 0.1295$, as we will be restricting our simulated contingency tables to the dimensions we state below.

To follow exactly the footprints of \citet{BS}, we consider $M=10^4$ replicates of contingency tables with $I=5$ rows and $J=8$ columns, containing $n=100$ observations. For each of the methods based on permutations, we chose $B=999$ as the number of resamples and we use the algorithm by \citet{Patefield} to uniformly draw the contingency tables with given marginals.

\begin{figure}%[!p]
	\centering\includegraphics[width=.7\textwidth]{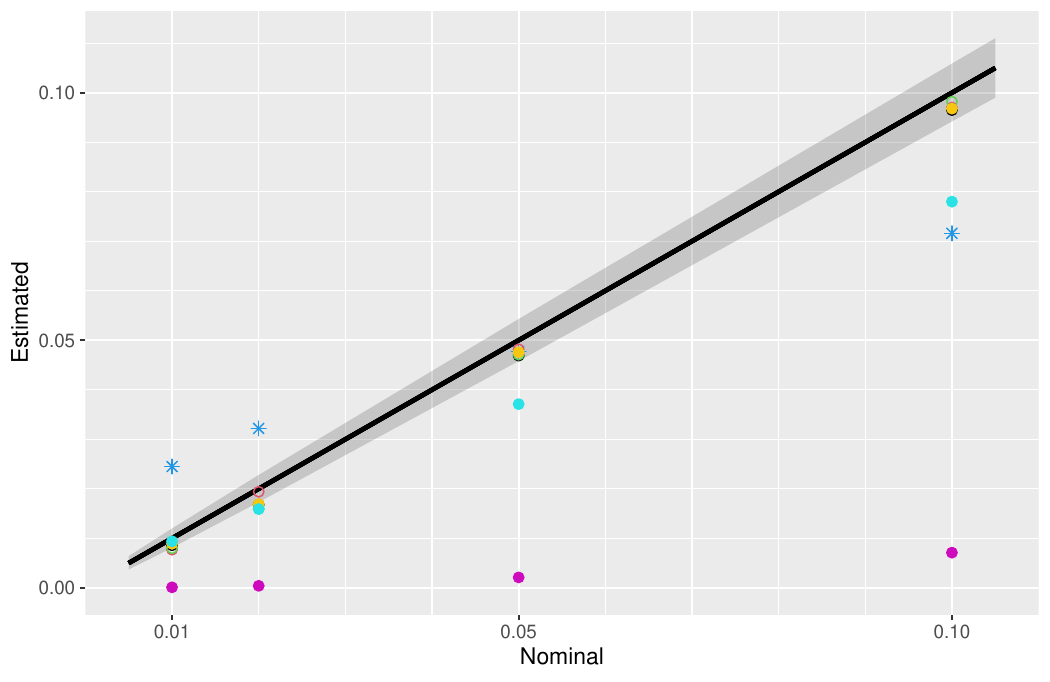}
	\caption{Empirical power under the null hypothesis ($\hat\alpha$) versus nominal significance level ($\alpha$), for the decaying marginals model, comparing our distance covariance method (golden points), Pearson's chi-squared test (pale blue), Pearson's test with permutations (dark red), the USP (black), the USP with the asymptotic approximation (turquoise), Fisher's exact test (green) and the $G$-test (purple). The gray shadow is a 95 \% confidence band for $\hat\alpha$ given $\alpha$.}
	\label{Calibr}
\end{figure}

For $\eps=0$ we can see how we calibrate significance. Figure~\ref{Calibr} shows the results with our method for some reference values of nominal $\alpha$, and allows for a comparison with competing techniques. We see that we control type I error very satisfactorily, both when considering our results only and when comparing them with Pearson's test with permutations, the USP and Fisher's exact test. All the aforementioned tests perform satisfactorily in terms of calibration of $\alpha$. The $G$-test, however, proves to be far too conservative. Pearson's chi-squared fails, too, when it comes to controlling the type I error, but does so in a less dramatic fashion (and it actually produces a good result for nominal $\alpha$ of 0.05). To find an explanation to this phenomenon, one should note that the model we are using features very small expected cell counts, which will tend to break down the heuristic rules as to when to use the chi-squared distribution with $(I-1)(J-1)$ degrees of freedom to compute $p$-values or not.

\begin{figure}%[!p]
	\centering\includegraphics[width=.7\textwidth]{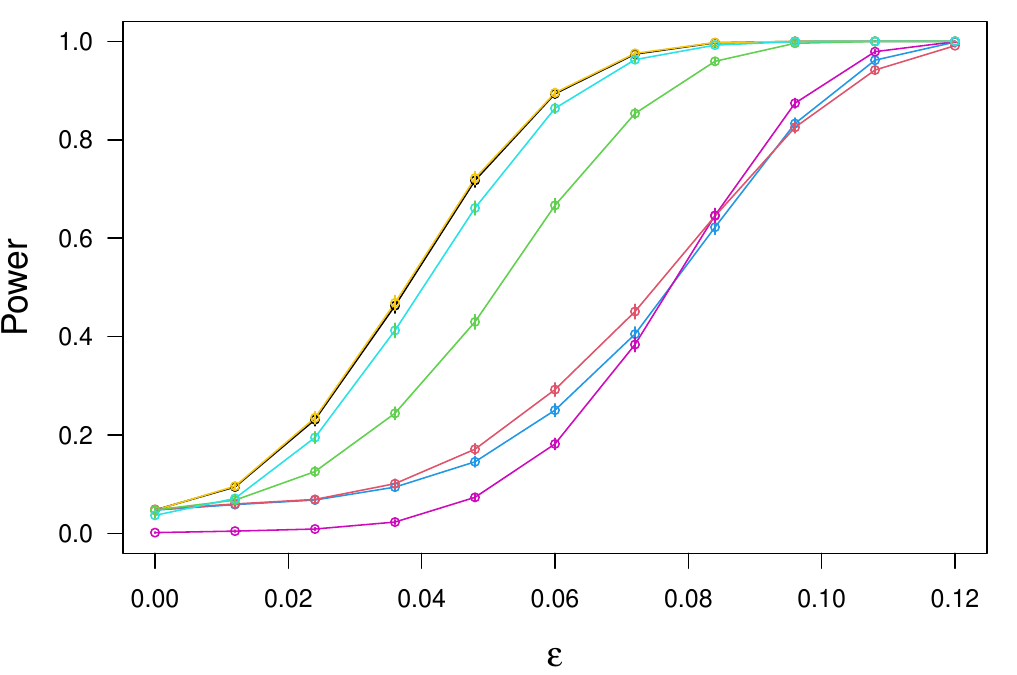}
	\caption{Power curve comparison for the decaying marginals model, displaying our distance covariance method (golden curve), Pearson's chi-squared test (pale blue), Pearson's test with permutations (dark red), the USP (black), the USP with asymptotic approximation (turquoise), Fisher's exact test (green) and the $G$-test (purple). The $5\times 8$ cells of each contingency table were filled with $n=100$ observations. $M=10^4$ replicates were considered. Error bars span from $-3$ to $+3$ standard deviations for each value of parameter $\eps$, which indicates the distance from the null hypothesis.}
	\label{Power_comparison_plot}
\end{figure}

In terms of power, Figure~\ref{Power_comparison_plot} shows that we perform very similarly to the USP (which shows how our derivation of the null distribution is correct and that the asymptotic approximation is not very far off when $n=100$). The power curve of Fisher's exact test is clearly under ours, whereas the one for the remaining classical methods is quite low for most values of $\eps$.

Other than the theoretical insight that using distance covariance provides (i.e., characterising general independence, the relationship to kernels and global tests, and so forth), we provide a relevant practical improvement with respect to the USP --- running time. Our experiments show that we are 3 orders of magnitude faster in testing than the USP. This remarkable difference in speed is not due to anything being intrinsically slow about computing the USP statistic, but it is simply a consequence of comparing a testing approach that uses a closed-form null distribution with another one that requires almost a thousand permutations in its default settings \citep{BS}. All this refers to the USP as provided by its original authors. When resorting to its asymptotic null distribution, as the one we are presenting for the distance covariance, no relevant discrepancy in computation times between the two methods is observed.

\subsection{Energy-distance test of goodness of fit}\label{simu:ed}

We will first summarize the notion of Hardy--Weinberg equilibrium (HWE), an important genetic concept that was independently introduced in 1908 by the eponymous authors \citep{Hardy,Weinberg}. Let us consider a biallelic locus, whose alleles we will denote as $A_1$ and $A_2$. Under panmixia and in the absence of evolutionary influences, the frequencies of both alleles and of each possible genotype ($A_1A_1$, $A_1A_2$ and $A_2A_2$) remain constant from generation to generation. If we use the following notation for the allele frequencies:
$$
\theta_1:=f(A_1);\;\;\; \theta_2:=f(A_2);
$$
the genotype frequencies that are to be maintained under the HWE are:
$$
f(A_1A_1)=\theta_1^2;\;\;\; f(A_1A_2)=2\theta_1\theta_2;\;\;\; f(A_2A_2)=\theta_2^2;
$$
where $\theta_1+\theta_2=1$. We point out that the \emph{frequencies} that geneticists denote by $f$ are what a statistician would call \emph{proportions} in the population. It is also noteworthy that those frequencies that the HWE predicts are the terms of the expansion of
$$\left( \theta_1+\theta_2\right)^2 $$
as a sum.

\begin{figure}[!htbp]
	\centering\includegraphics[width=.9\textwidth]{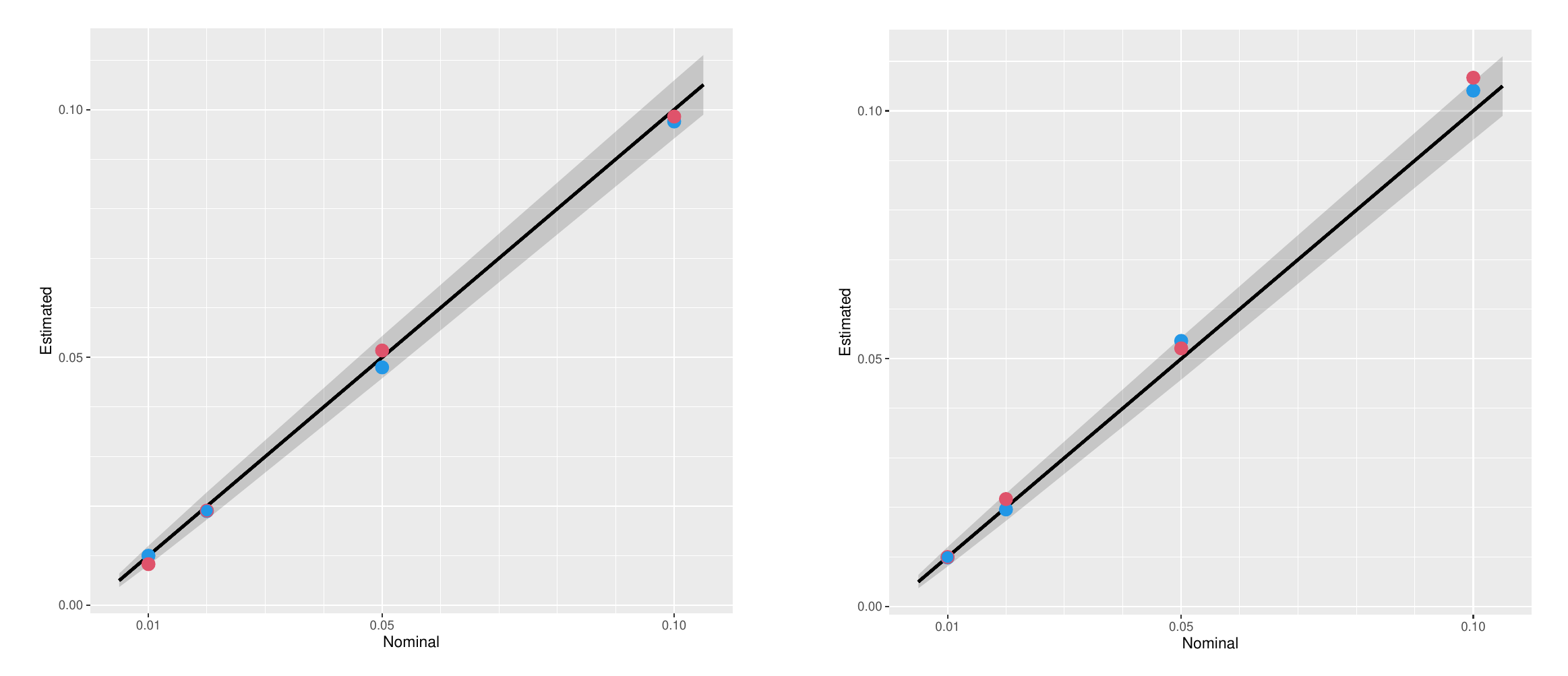}
	\caption{Empirical power under the null hypothesis ($\hat\alpha$) versus nominal significance level ($\alpha$), for the goodness-of-fit test of the biallelic Hardy--Weinberg equilibrium, when $\theta_1=\frac{2}{3}$ (left-hand plot) and $\theta_1=\frac{1}{2}$ (right). Red dots correspond to our energy distance method; blue are those for Pearson's chi-squared test. The gray shadow is a 95 \% confidence band for $\hat\alpha$ given $\alpha$.}
	\label{calibr:gof2}
\end{figure}

We will now start the simulations by showing the calibration of significance for some reference values of nominal $\alpha$ for our energy-distance test and the chi-squared test of goodness of fit. Based on the values for the allele frequencies we have encountered in the real data examples that we will be presenting in Subsection~\ref{rdata:ed}, we have chosen $\frac{2}{3}$ and $\frac{1}{2}$ as representative values of $\theta_1$ for our simulations. Figure~\ref{calibr:gof2} shows that both our method and the $\chi^2$ test perform well in terms of type I error. Every simulation in this subsection will take $n=500$ observations for each of the $M=10^4$ replicates. The sample size is a rounding of the one we have in Section~\ref{rdata:categorical}, but our numerical experiments show qualitatively similar conclusions for other values of $n$.

We now introduce two models that depart from the null hypothesis. For model 2S, we first consider the HWE genotype frequencies for the case where $\theta_1=\frac{2}{3}$:

\begin{center}
	\begin{tabular}{ccc}
		$A_1A_1$& $A_1A_2$ & $A_2A_2$  \\
		\hline
		$\frac{4}{9}$& $\frac{4}{9}$ &$\frac{1}{9}$  \\
	\end{tabular}
\end{center}

And we introduce a parameter $s\in[0,1]$ which is zero under the null hypothesis and it increases as so does the distance to $H_0$:
\begin{center}
	\begin{tabular}{ccc}
		$A_1A_1$& $A_1A_2$ & $A_2A_2$  \\
		\hline
		$\frac{4(1-s)}{9}$& $\frac{4(1-s)}{9}$ &$\frac{1+8s}{9}$  \\
	\end{tabular}
\end{center}

On the other hand, model 2K introduces parameter $k\in[0,1]$, which increases as so does the divergence from HWE with $\theta_1=\theta_2=\frac{1}{2}$:
\begin{center}
	\begin{tabular}{ccc}
		$A_1A_1$& $A_1A_2$ & $A_2A_2$  \\
		\hline
		$\frac{1-k}{4}$& $\frac{k+1}{2}$ &$\frac{1-k}{4}$  \\
	\end{tabular}
\end{center}

\begin{figure}[!htbp]
	\centering\includegraphics[width=.9\textwidth]{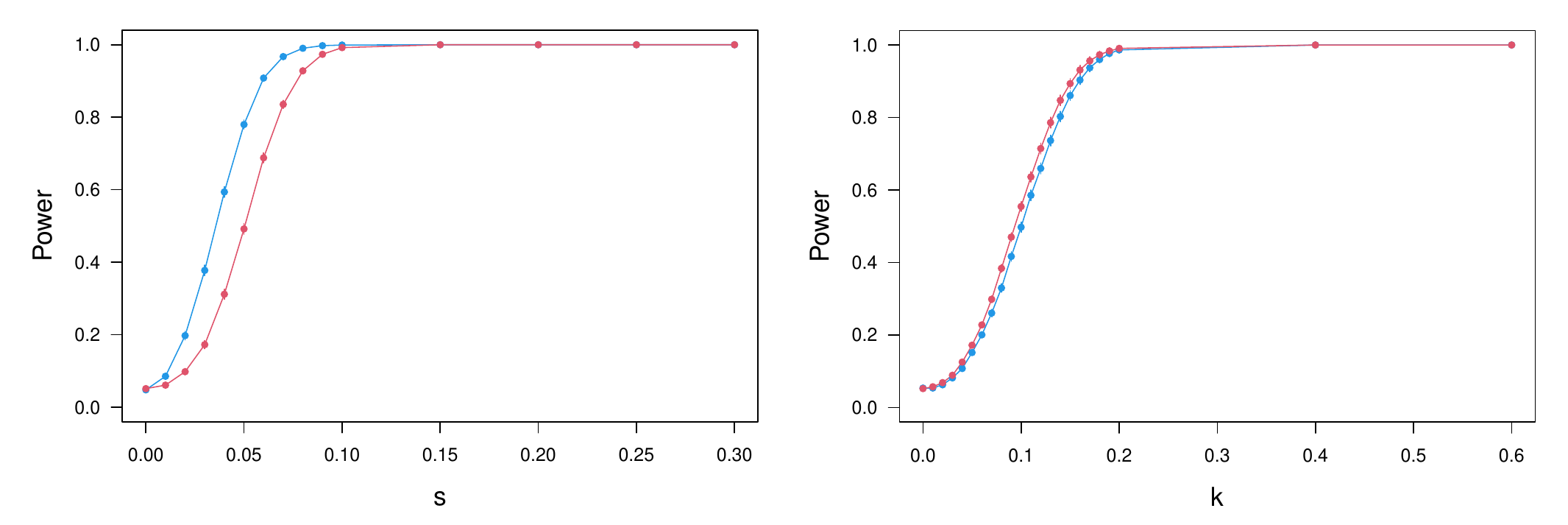}
	\caption{Power curve comparison for models 2S (left) and 2K (right), displaying our energy distance method (red lines and dots) and Pearson's chi-squared test (blue). $M=10^4$ replicates with sample size $n=500$ were considered. Error bars are barely visible in this case, but they span from $-3$ to $+3$ standard deviations for each value of parameters $s$ and $k$, which in turn indicate the distance to the null hypothesis.}
	\label{power:2allele}
\end{figure}

We present power curves for models 2S and 2K for both $\mathcal E$ and the $\chi^2$ test in Figure~\ref{power:2allele}. We observe that both tests perform very satisfactorily, even for divergences from the null hypothesis that are not the highest in magnitude.

In order not to restrict ourselves to the case where the number of categories is only $3$, we will now generalize the notion of HWE. One way of doing so would be to increase the ploidy, which would yield as genotype frequencies the terms of the binomial expansion of
$$\left( \theta_1+\theta_2\right)^c $$
for $c>2$. We will however opt for a generalisation that one can encounter in humans, that is, increasing the number of possible alleles. Let us consider a triallelic locus with allele frequencies
$$
\theta_1:=f(A_1);\;\;\; \theta_2:=f(A_2);\;\;\; \theta_3:=f(A_3);
$$
where $\theta_1+\theta_2+\theta_3=1$. Then the Hardy--Weinberg genotype frequencies are:
\begin{center}
	\begin{tabular}{cccccc}
		$A_1A_1$& $A_2A_2$ & $A_3A_3$& $A_1A_2$& $A_1A_3$ & $A_2A_3$  \\
		\hline
		$\theta_1^2$&$\theta_2^2$&$\theta_3^2$& $2\theta_1\theta_2$ &$2\theta_1\theta_3$ &$2\theta_2\theta_3$   \\
	\end{tabular}
\end{center}

As with the biallelic case, we first consider a scenario where the allele frequencies are unbalanced: $\theta_1=0.70$, $\theta_2=0.25$ and $\theta_3=0.05$. Model 3S departs from the HWE for those values as parameter $s\in[0,1]$ increases within its range:
\begin{center}
	\begin{tabular}{cccccc}
		$A_1A_1$& $A_2A_2$ & $A_3A_3$& $A_1A_2$& $A_1A_3$ & $A_2A_3$  \\
		\hline
		$0.49(1-s)$&$\frac{1+15s}{16}$&$0.0025(1-s)$& $0.35(1-s)$ &$0.07(1-s)$ &$0.025(1-s)$   \\
	\end{tabular}
\end{center}

And we also consider the case where $\theta_1=\theta_2=\theta_3=\frac{1}{3}$. By introducing parameter $k\in[0,1]$ to tune the intensity of the departure from the null, we define model 3K:
\begin{center}
	\begin{tabular}{cccccc}
		$A_1A_1$& $A_2A_2$ & $A_3A_3$& $A_1A_2$& $A_1A_3$ & $A_2A_3$  \\
		\hline
		$\frac{2k+1}{9}$& $\frac{2k+1}{9}$& $\frac{2k+1}{9}$ &$\frac{2-2k}{9}$ &$\frac{2-2k}{9}$ &$\frac{2-2k}{9}$   \\
	\end{tabular}
\end{center}

\begin{figure}[!htbp]
	\centering\includegraphics[width=.9\textwidth]{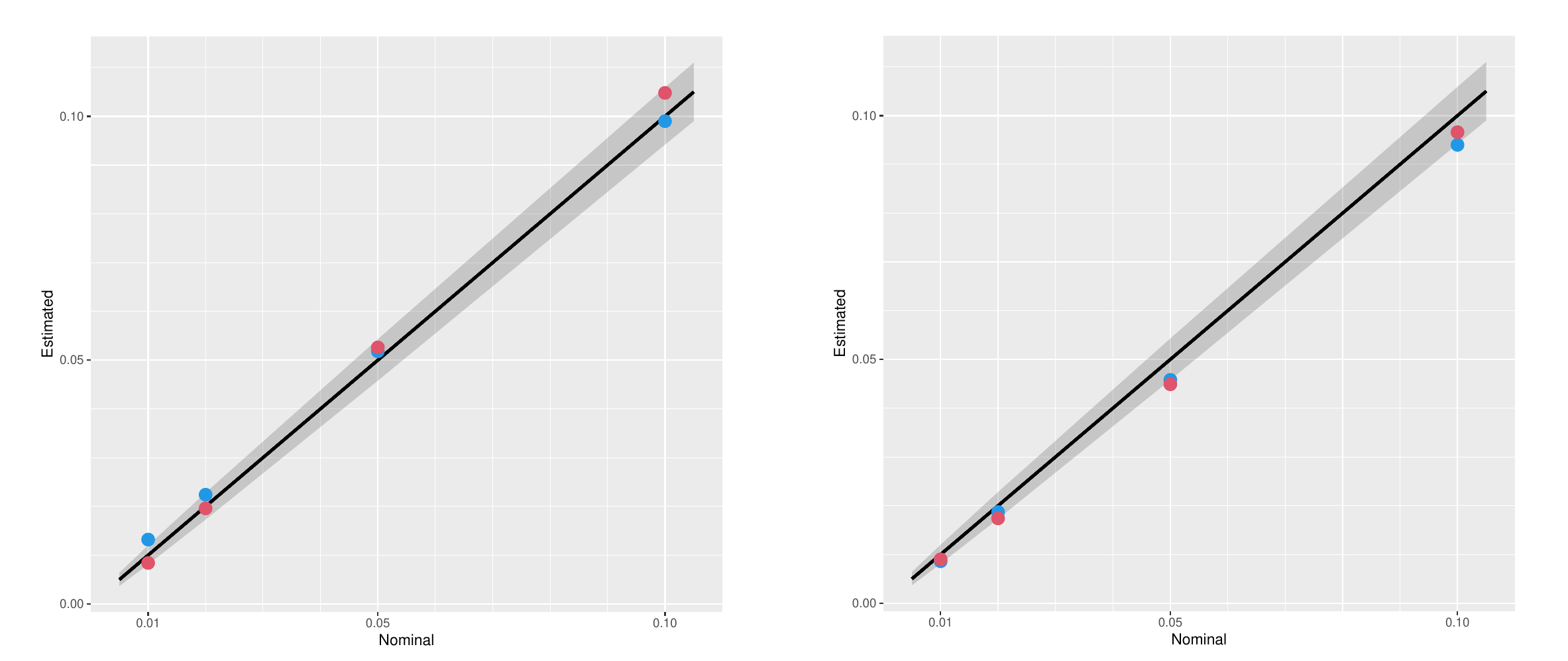}
	\caption{Empirical power under the null hypothesis ($\hat\alpha$) versus nominal significance level ($\alpha$), for the goodness-of-fit test of the triallelic Hardy--Weinberg equilibrium, when $(\theta_1,\theta_2, \theta_3)=(0.70, 0.25, 0.05)$ (left-hand plot) and $\theta_1=\theta_2=\theta_3=\frac{1}{3}$ (right). Red dots correspond to our energy distance method; blue are those for Pearson's chi-squared test. The grey shadow is a 95 \% confidence band for $\hat\alpha$ given $\alpha$.}
	\label{calibr:gof3}
\end{figure}

\begin{figure}[!htbp]
	\centering\includegraphics[width=.9\textwidth]{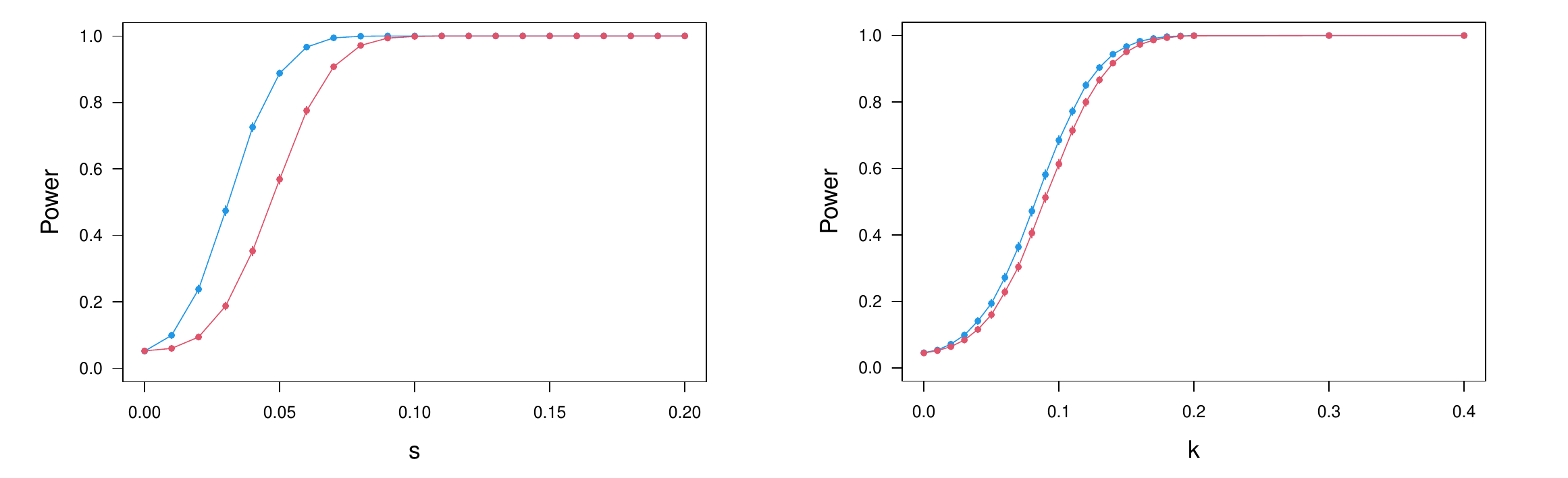}
	\caption{Power curve comparison for models 3S (left) and 3K (right), displaying our energy distance method (red lines and dots) and Pearson's chi-squared test (blue). $M=10^4$ replicates with sample size $n=500$ were considered. Error bars are barely visible in this case, but they span from $-3$ to $+3$ standard deviations for each value of parameters $s$ and $k$, which in turn indicate the distance from the null hypothesis.}
	\label{power:3allele}
\end{figure}

Figure~\ref{calibr:gof3} shows that, once again, both the energy distance and Pearson's chi-squared control type I error. The power curves in Figure~\ref{power:3allele} show $\mathcal E$ a bit below the $\chi^2$, but we do not perform a great deal worse.

\section{Real data analyses}\label{rdata:categorical}

To complete the numerical analyses in Section~\ref{simu:categorical}, we now demonstrate the applicability of the methodology introduced in this paper. We introduce two examples of interest to biomedical practice that arise from a data set produced by us \citep{Facal:Scand}. Subsection~\ref{rdata:dc} explores the potential of our distance-covariance independence test for interpreting the clinical significance of polygenic scores, whereas Subsection~\ref{rdata:ed} presents real-life examples of the Hardy--Weinberg models introduced in Subsection~\ref{simu:ed}.

\subsection{Distance-covariance test of independence}\label{rdata:dc}

We begin by showing with a real biomedical example how our test for dependence can be used in practice. We consider data from \citet{Facal:Scand}, where we observe $n=427$ patients of schizophrenia. For each of them, we consider a categorical variable $X$ indicating how chronic the psychiatric disorder is in that person (an index with four possible values, based on the admission history in health facilities), and another categorical variable $Y$ which indicates the PRS tercile (i.e., whether the \emph{polygenic risk score} for schizophrenia of the patient is low, medium or high).

Although the clinical utility of PRSs is very limited at the individual level, they may be useful for the identification of specific quantiles of risk for stratification of a population to apply specific interventions \citep{PRS:Topol}. This is why it makes the most sense to consider PRS as a categorical variable (and not one with many categories) instead of working with its raw individual scores. The data for our example can be seen in Table~\ref{tab:chronic}.

\begin{table}%[!p]
	\centering
	%	\begin{measuredfigure}
		\caption{Contingency table for the chronicity data set.}
		\begin{tabular}{cccc|c}
			\multicolumn{1}{c|}{Chr. \textbackslash $\:$PRS} & $\mathrm{T}_1$     & $\mathrm{T}_2$     & $\mathrm{T}_3$     &  \\
			\cmidrule{1-4}    \multicolumn{1}{c|}{Low} & $12$ & $9$ & $4$ & $25$ \\
			\multicolumn{1}{c|}{Middle-Low} & $37$ & $20$ & $29$ & $86$ \\
			\multicolumn{1}{c|}{Middle-high} & $40$ & $58$ & $44$ & $142$ \\
			\multicolumn{1}{c|}{High} & $53$ & $55$ & $66$ & $174$ \\
			\midrule
			& $142$   & $142$   & $143$ & $427$  \\
		\end{tabular}%
		%	\end{measuredfigure}
	\label{tab:chronic}
\end{table}

We can now apply the different methods of Section~\ref{simu:categorical} to our data set. Pearson's test yields similar results with and without permutations, due to the lack of low (expected) cell counts. In both cases, the $p$-value is around 0.025 and one would reject independence for a nominal $\alpha$ of 0.05. The $G$-test offers a $p$ of 0.022, in line with Pearson's. Fisher's exact test also does not diverge much, with 0.024. Finally, the USP and the distance covariance yield $p$-values of 0.046 and 0.045. All things considered, in this case one would tend to reject the null hypothesis of independence (when $\alpha=0.05$), which is consistent with the hypothesis that the PRS can measure how ``sick'' a patient is (or, more generally, how intense the trait of interest is).

\subsection{Energy-distance test of goodness of fit}\label{rdata:ed}
We will now see two examples of how one can test for goodness of fit with our methodology. Let us consider again the cohort of $n=427$ individuals by \citet{Facal:Scand}. As previously mentioned, a frequent quality control for GWAS data is whether or not each SNP is in HWE in the control group.

We will firstly focus the biallelic SNP rs9545047 because it is one of the variants in the most current list of loci known to influence gene expression in relationship with schizophrenia, as per Extended Data Table 1 in \citet{PGC3}. This SNP has also the peculiarity of not being in a protein-coding gene, but in one that is transcribed into long intergenic nonprotein coding RNA (lincRNA). For this locus, we observe genotype \textit{AA} 139 times; \textit{CA}, 232 times and \textit{CC}, 56 times. Using the online tool UCSC Genome Browser \citep{UCSC}, we can retrieve some useful information about this SNP, including the allele frequencies according to the GnomAD database, which gives us: $$f(C)\approx0.41\,.$$

GnomAD v4.1.0 offers allele frequencies for different ancestries, and we have chosen the value for European (non-Finnish) population, since it is the best match for the geographical origin of our 427 individuals, which are from the northwestern Iberian peninsula. We have opted for GnomAD because it is the online resource for human population genetics with the largest sample size that we are aware of.

Therefore, the expected cell counts are:
\begin{center}
	\begin{tabular}{ccc}
		$AA$& $CA$ & $CC$  \\
		\hline
		$148.6$& $206.6$ &$71.8$  \\
	\end{tabular}
\end{center}

When applying our energy testing procedure, it yields a $p$-value of $0.027$, which coincides with the one obtained with Pearson's. This means that both tests would reject the null hypothesis for conventional nominal values of $\alpha$ like $0.05$. This is a perfectly logical result for a SNP linked to schizophrenia, which is expected to have one of its haplotypes at a frequency that departs from the one that would be encountered under the HWE. One should also note that SNPs like this one are not left out during the quality control phase of the GWAS because the Hardy--Weinberg filter only applies to the control group (in our case, a pool of individuals not presenting schizophrenia).

Given that there are not many triallelic SNPs, we will just be considering one of them for illustrative purposes, without giving much profound interpretation to the results. We choose SNP rs2594292, for which the observed genotypes are:
\begin{center}
	\begin{tabular}{cccccc}
		$AA$& $GG$ & $TT$ & $AG$ & $AT$ & $TG$  \\
		\hline
		$214$& $34$ &$0$ & $148$ & $16$ & $15$  \\
	\end{tabular}
\end{center}

Once again resorting to GnomAD, we get the following population allele frequencies:
$$f(A)\approx0.69;\;\;f(G)\approx0.26;\;\;f(T)\approx0.05\,.$$
Using them to calculate the expected cell counts, we get a $p$-value of $0.24$ with our method and of $0.07$ with Pearson's. In this case we observe more dissimilar results, but with none of the tests finding significant evidence of divergence from the HWE with nominal $\alpha$ of $0.05$, which is a logical result for any SNP not known to be linked to schizophrenia.

\section{Discussion and Conclusion}

We have proposed a new test for the independence of categorical variables (one of the most often tested hypotheses in biomedical research) by using distance covariance, an association measure that characterizes general statistical independence. As we allow for arbitrary dimensions of the contingency table, this extends the possibilities we showed on previous work \citep{Epistasis:paper} for the $3\times3$ case. We have also developed a novel testing strategy for the goodness of fit to a discrete distribution. For both methods, we demonstrate good performance and applicability, with simulations and analyses of relevant biomedical examples.

The test statistic we derive for independence happens to have a simple algebraic expression similar in spirit to that of Pearson's $\chi^2$ test. We are not the first to see the connection between the two tests, as it was already mentioned in Remark 3.12 of \citet{Lyons} and explored in some detail in the final section of \cite{DJ}. Nevertheless, the proofs we provide are original and we are the first ones (to our knowledge) to analyse the matter in detail. On top of that, we are not aware of any previous instance in the literature where a test for goodness of fit to a discrete distribution is built based on energy statistics.

Another test for independence that is related to ours is the one in \citet{BS}, initially introduced in \citet{BKS}. The main conceptual difference in our approaches is that we derive the asymptotic null distribution of our $V$-statistic and are able to satisfactorily use it in practice, whereas their testing is based on permutations (of a $U$-statistic). It is also noteworthy that, in that article, no mention is made of distance--based association measures, a relationship that we thoroughly explore. In return, we obtain from their results the conclusion that our test statistic is very close to being the minimum-variance unbiased estimator of the population USP-divergence statistic. As they indicate, if one assumes that the population quantity is meaningful (and we now know it is, given its connection to distance covariance), then the test statistic is a very good estimator of it.

A remarkable pragmatical difference between our goodness-of-fit test and the one for independence is that the former does not require to plug in any frequencies to then estimate the multinomial covariance matrix and get the coefficients of the linear combination of chi-squared's. In this case, the $p_i$'s are fixed and known, since they are given by the null hypothesis. However, when testing whether or not the population distribution belongs to a certain family of distributions, one would need to plug in the parameters in which the family is indexed. The effect that the estimation of such parameters has in $U-$ and $V-$statistics has been studied by authors such as \citet{Wet:Randles} and \citet{Gamero}.

All in all, we have presented new methodology to address important problems of practitioners, proven solid theoretical properties, explored connections with well-known methods, and illustrated all of it in simulated and real data sets. Future and current lines of work include extending these techniques to the study of associations between categorical and continuous data \citep{GWAS:paper}.

\section*{Acknowledgements}

This work has been supported by project \mbox{PID2020-116587GB-I00}, granted by MICIU/AEI/10.13039/ 501100011033 (Spanish Ministry of Science). FCP's research was carried out under the \mbox{FPU19/04091} grant of the Spanish Ministry of Universities. The schizophrenia data set  was generated under support of the Instituto de Salud Carlos III (grant number ISCIII/PI14/01020) to Javier Costas, cofounded by European Regional Development Fund (ERDF). The authors are grateful to Prof Jelle J Goeman (Leiden University Medical Center) and Prof Thomas Berrett (University of Warwick) for fruitful discussions on the topic of this article. FCP and DE would also like to show their appreciation to all who made the CEN-IBS 2023 conference happen, since it was an invaluable chance to present an earlier version of this work and to network with colleagues.

\newpage

%\vspace*{2cm}

\renewcommand{\thesection}{Appendix \Alph{section}:}
\setcounter{section}{0}
\section{Proof of Theorem~\ref{th:indep}}\label{proof:dist}
We will firstly show that the distance covariance test statistic has the compact form similar to Pearson's that we stated in the main manuscript, to then prove the asymptotic null distribution.

We will investigate the terms $\widehat{T}_1, \widehat{T}_2, \widehat{T}_3$ one by one, to then see how $\widehat V$ can be written as a simple expression.
\begin{align*}
	\widehat{T}_1 &= \frac{1}{n^2} \sum_{l,m =1}^n d(X_l,X_m) \, d(Y_l,Y_m) \\
	&= \frac{1}{n^2} \sum_{l,m =1}^n 1_{\{X_l \neq X_m, Y_l \neq Y_m\}} \\
	&= \frac{1}{n^2} \sum_{l,m =1}^n \left( 1 - 1_{ \{X_l = X_m\} } - 1_{ \{Y_l = Y_m\} } + 1_{\{X_l = X_m, Y_l = Y_m\}}\right)  \\
	&= 1 - \frac{1}{n^2} \sum_{i=1}^{I} n_{i \cdot}^2 - \frac{1}{n^2} \sum_{j=1}^{J} n_{\cdot j}^2 +  \frac{1}{n^2}  \sum_{i=1}^{I} \sum_{j=1}^{J} n_{ij}^2.
\end{align*}  
For $\widehat{T}_2$, we first observe that 
$$
\sum_{m =1}^n d(X_l,X_m)
= \sum_{m =1}^n \left( 1 - 1_{ \{X_l = X_m\} } \right) 
= n - n_{X_l \cdot}
$$
and hence 
\begin{align*}
	\widehat{T}_2 &= \frac{1}{n^3} \sum_{l=1}^n (n- n_{X_l \cdot}) (n- n_{\cdot Y_l}) \\
	&= \frac{1}{n^3} \sum_{i=1}^{I} \sum_{j=1}^{J} (n - n_{i \cdot}) (n - n_{\cdot j }) n_{ij} \\
	&= 1 - \frac{1}{n^2} \sum_{i=1}^{I} n_{i \cdot}^2 - \frac{1}{n^2} \sum_{j=1}^{J} n_{\cdot j}^2 +   \frac{1}{n^3} \sum_{i=1}^{I} \sum_{j=1}^{J} n_{i \cdot} n_{\cdot j} n_{ij}.
\end{align*}	
Finally,
$$
\sum_{l,m =1}^n d(X_l,X_m) = \sum_{l =1}^n \left( n - n_{X_l \cdot} \right)   = n^2 - \sum_{i=1}^I n_{i \cdot}^2
$$
and hence
\begin{align*}
	\widehat{T}_3 &= \frac{1}{n^4} \left( n^2 - \sum_{i=1}^I n_{i \cdot}^2\right)  \left( n^2 - \sum_{j=1}^J n_{\cdot j}^2\right) \\
	&= 1 - \frac{1}{n^2} \sum_{i=1}^{I} n_{i \cdot}^2 - \frac{1}{n^2} \sum_{j=1}^{J} n_{\cdot j}^2 +  \frac{1}{n^4} \sum_{i=1}^{I} \sum_{j=1}^{J} n_{i \cdot}^2 n_{\cdot j}^2.
\end{align*}

When adding up the terms to obtain $\widehat{V}$, the terms $1$, $\frac{1}{n^2} \sum_{i=1}^{I} n_{i \cdot}^2$ and $\frac{1}{n^2} \sum_{j=1}^{J} n_{\cdot j}^2$ all cancel out and we obtain
\begin{align*}
	\widehat{V}  &= \frac{1}{n^2}  \sum_{i=1}^{I} \sum_{j=1}^{J} n_{ij}^2 - \frac{2}{n^3} \sum_{i=1}^{I} \sum_{j=1}^{J} n_{i \cdot} n_{\cdot j} n_{ij} + \frac{1}{n^4} \sum_{i=1}^{I} \sum_{j=1}^{J} n_{i \cdot}^2 n_{\cdot j}^2 \\
	&= \frac{1}{n^2}  \sum_{i=1}^{I} \sum_{j=1}^{J} \left( n_{ij} - \frac{1}{n} n_{i \cdot} n_{\cdot j}\right) ^2 \\
	&= \frac{1}{n^2}  \sum_{i=1}^{I} \sum_{j=1}^{J}  (n_{ij} - n^*_{ij})^2,
\end{align*} 	
which is what we wanted to achieve.

Now, to start the way towards the asymptotic null distribution, let $\spz$ be either $\{1,\ldots,I\}$ or $\{1,\ldots,J\}$. Then the discrete metric on $\spz$
\[
d(z,z')=1-\delta_{zz'},
\]
is dual to the following kernel in the sense of \citet{Sejdinovic}:
\[
k(z,z')=\delta_{zz'},
\]
which is known as the \emph{discrete kernel}. Then clearly one can take the dummy function on each of $\spx$ and $\spy$ as a feature map of the corresponding kernel/distance. We will denote them by $\phi:\spx\longrightarrow\R^I$ and $\psi:\spy\longrightarrow\R^J$, where:
$$
\phi_i(X) = 1_{\{X=i\}} , \quad \psi_j(Y) = 1_{\{Y=j\}}.
$$
Now we construct matrices $\bU=(U_{ij})_{n\times I}$ and $\bV=(V_{ij})_{n\times J}$ by transforming the $X$ and $Y$ samples with the feature maps: 
$$
U_{ki} = \phi_i(X_k) \quad V_{kj} = \psi_j(Y_k) .
$$
Note that each of row of the previous matrices contains an observation of $\phi(X)\sim\Multib(\bq)$ or $\psi(Y)\sim\Multib(\br)$ (respectively). Therefore:
$$
\bone\transp\bU\sim\Multin_I(n,\bq)
$$
$$
\bone\transp\bV\sim\Multin_J(n,\br)
$$
Now, applying Equation 3 in \citet{DJ} to our feature maps, we get:
$$
n \, \dCovh_{\text{discrete}}^2(X,Y)=\frac{1}{n}\sum_{i=1}^I\sum_{j=1}^J[\bU\transp(\bI_n-\bH)\bV]^2_{ij} ,
$$
where $\bI_n$ is the $n\times n$ identity matrix and $\bH=\frac{1}{n}\bone\bone\transp$ has constant entries equal to $\frac{1}{n}$. If we now define $\bC\equiv(C_{ij})_{I\times J}:=\frac{1}{\sqrt n}\bU\transp(\bI_n-\bH)\bV$, we can compactly write our test statistic as a trace:
$$
n \, \dCovh_{\text{discrete}}^2(X,Y)=\tr[\bC\bC\transp]=\tr[\bC\transp\bC]=\sum_{i=1}^I\sum_{j=1}^JC_{ij}^2 .
$$
Expressing an empirical distance covariance as a trace of a matrix product, as we did above, is not unusual \citep{TEOD} and indeed it is a very computationally efficient way of evaluating it. Nonetheless, for continuing the proof we are going to write:
$$
n \, \dCovh_{\text{discrete}}^2(X,Y)=\bc\transp \bc ;
$$
where $\bc:=\vecop(\bC)\in\R^{IJ}$ is the vectorization of matrix $\bC$ (i.e., its image by the linear isomorphism $\R^{I\times J}\cong \R^{IJ}$).

If one adds a vector with constant components $\ba=a\bone$ to a column or row of a matrix, the result of centering it with matrix $\bI-\bH$ will be the same. Therefore, we can expand $\bC$ as:
$$
\bC=
\frac{1}{\sqrt{n}} (\bU\transp-\bq\bone\transp)(\bI-\bH)(\bV-\bone\br\transp)=
$$
$$
=\frac{1}{\sqrt{n}} (\bU\transp-\bq\bone\transp)(\bV-\bone\br\transp)-
\frac{1}{n^{3/2}}(\bU\transp-\bq\bone\transp)\bone\bone\transp(\bV-\bone\br\transp).
$$
The second term of the previous sum is:
$$
\bD:=\frac{1}{\sqrt n} \:
\left[ \frac{1}{\sqrt n}\begin{pmatrix}
	\sum_{m=1}^n \left( \phi_1(X_m)-q_1\right)  \\
	\ldots \\
	\sum_{m=1}^n \left( \phi_I(X_m)-q_I\right)
\end{pmatrix}\right] \:
\left[ \frac{1}{\sqrt n}
\left( 
\sum_{m=1}^n \left( \psi_1(Y_m)-r_1\right),
\ldots,
\sum_{m=1}^n \left( \psi_J(Y_m)-q_J\right)
\right) 
\right]
$$
By the central limit theorem, it is easy to see that each entry $D_{ij}$ of $\bD$ converges in probability to zero, owing to the fact that:
$$
\frac{1}{\sqrt n}\sum_{m=1}^{n}\left( \phi(X_m)-\bq\right) \distrilim\Normal_I(\bzero,\bA)
$$
$$
\frac{1}{\sqrt n}\sum_{m=1}^{n}\left( \psi(Y_m)-\br\right) \distrilim\Normal_J(\bzero,\bB) .
$$
Hence, $\vecop(\bD)$ converges in probability to the $IJ-$dimensional null vector, and the limit in distribution of $\bc$ will be that of the vectorisation of:
$$\bE:=\frac{1}{\sqrt{n}} (\bU\transp-\bq\bone\transp)(\bV-\bone\br\transp).$$
We can write the $(i,j)$th entry of the previous matrix as: $E_{ij}=\frac{1}{\sqrt n}\sum_{m=1}^{n} G_{mij}$, where
$$
G_{mij}=\left( \phi_i(X_m)-q_i\right) \left( \psi_j(Y_m)-r_j\right).
$$
Now, we see that we can apply the CLT to
$$
\vecop(\bE)=\frac{1}{\sqrt{n}}\sum_{m=1}^{n}\vecop(\bG_m).
$$
For fixed $m\in\{1,\ldots,n\}$, let us see how the first and second moments of $\vecop(\bG)\equiv\vecop(\bG_m)$ look like. For $i\in\{1,\ldots,IJ\}$, the $i$th component of $\E[\vecop(\bG)]$ vanishes under the null hypothesis (i.e., independence of $X$ and $Y$):
$$
\E[G_{(i-1)\%I+1,\ceil{i/I}}]
=\E[\left( \phi_{(i-1)\%I+1}(X)-q_{(i-1)\%I+1}\right)  ]
 \E[\left( \psi_{\ceil{i/I}}(Y)-r_{\ceil{i/I}}\right)  ]=0\cdot0=0.
$$
We have used the notation $\%$ to indicate the remainder of an integer division, and $\ceil{\cdot}$ for the ceiling.

The $(i,j)$th entry of the variance-covariance matrix of $\vecop(\bG)$ is:
$$
\Cov(G_{(i-1)\%I+1,\ceil{i/I}},G_{(j-1)\%J+1,\ceil{j/J}})
=
$$
$$
=
\E[\left( \phi_{(i-1)\%I+1}(X)-q_{(i-1)\%I+1}\right) 
\left( \phi_{(j-1)\%J+1}(X)-q_{(j-1)\%J+1}\right)
 ]$$
 $$\times
\E[\left( \psi_{\ceil{i/I}}(Y)-r_{\ceil{i/I}}\right)
\left( \psi_{\ceil{j/J}}(Y)-r_{\ceil{j/J}}\right)
  ]=
  $$
  $$
  =
a_{(i-1)\%I+1,(j-1)\%J+1}\,b_{\ceil{i/I},\ceil{j/J}}
=
[\bB\tensor\bA]_{ij}\; ,
$$
with $\tensor$ denoting the Kronecker product.

Applying the central limit theorem once more, we get the limiting distribution of $\bc$:
$$
\bc\distrilim
\Normal_{IJ}(\bzero,\bGamma);\quad \bGamma=\bB\tensor\bA
$$
Now, one would be tempted to take $\bGamma$ to the $-\frac{1}{2}$ and standardize $\bc$, but the reality is that $\bGamma$ is never of full rank because $\bA$ and $\bB$ never are. So we are going to first take some sort of matrix root and then consider its inverse, instead of the other way round.

Let us write $\bGamma=\bM \bM\transp$, where $\bM\in\R^{IJ\times r}$ has rank $r:=\rank(\bGamma)\leq IJ$. If $\bM^+$ denotes the Moore--Penrose (pseudo)inverse of $\bM$, we can easily conclude that:
$$
\bw:=\bM^{+}\bc\distrilim
\Normal_{r}(\bzero,\bI)
$$
by taking into account that
$$\bM^+\bGamma(\bM^+)\transp=\bM^+\bM(\bM^+\bM)\transp=\bM^+\bM\bM^+\bM=\bM^+\bM=\bI_r,$$
with the last equality owing to the fact of $\bM$ having full column rank.

We can finally go back to the expression of the empirical distance covariance:
$$
n \, \dCovh_{\text{discrete}}^2(X,Y)=
\bw\transp\bGamma \bw .
$$
As $\bGamma$ is symmetric, we can diagonalize it with an orthogonal modal matrix $\bQ\in\R^{IJ\times IJ}$:
$$
\bGamma=\bQ\transp\bLambda\bQ,
$$
where $\bLambda\in\R^{IJ\times IJ}$ is a diagonal matrix and has the eigenvalues of $\bB\tensor\bA$ in its diagonal (which are the $IJ$ products of the eigenvalues $\{\lambda_i\}_i$ and $\{\mu_j\}_j$ of $\bA$ and $\bB$, respectively).
This allows us to conclude:
$$
n \, \dCovh_{\text{discrete}}^2(X,Y)\distrilim\sum_{i,j}\lambda_i\mu_{j} Z_{ij}^2 ,
$$
where $\brc{Z_{ij}}_{i,j}$ are IID standard Gaussian.\qed

\vspace*{2cm}

\section{Proof of Theorem~\ref{th:gof}}\label{proof:gof}

We will first derive the compact expression of $\mathcal E_n$. To that purpose, we firstly recall the definition of energy distance:
\begin{equation}\label{def:E:app}
\mathcal E_n=n\left[ \frac{2}{n}\sum_{l=1}^n\E d(x_l,X)-\E d(X,X') -\frac{1}{n^2}\sum_{l,m=1}^n d(x_l,x_m) \right];
\end{equation}
where all the notation so far is the same as in the main manuscript.

We firstly note that, for the discrete metric: $\E d(x_l,X)=\Prob\{X\neq x_l\}$. Summing over $l$ and multiplying by $\frac{2}{n}$:
$$
\frac{2}{n}\sum_{l=1}^n \E d(x_l,X)=\frac{2}{n}\sum_{l=1}^n (1-\Prob\{X= x_l\})=\sum_{i=1}^I \frac{n_i}{n} (1-p_i)=\sum_{i=1}^I \hat p_i (1-p_i);
$$
where $\hat p_i:=\frac{n_i}{n}$ is the estimated probability of category $i\in\{1,\ldots,I\}$ given the sample.

Secondly, we write the straightforward identity $$\E d(X,X')=1-\sum_{i=1}^I p_i^2.$$

And finally, for the remaining term of $\mathcal E_n/n$, we apply similar arguments to conclude:
$$
\frac{1}{n^2}\sum_{l,m=1}^n d(x_l,x_m)=1-\sum_{i=1}^I \hat p_i^2 .
$$

Now, adding up the three expressions:
$$
\frac{\mathcal E_n}{n}=2\sum_{i=1}^I \hat p_i(1-p_i)-\left[ 1-\sum_{i=1}^Ip_i^2\right] -\left[ 1-\sum_{i=1}^I\hat p_i^2\right] =
$$
$$
-2\sum_{i=1}^I\hat p_i p_i + \sum_{i=1}^I p_i^2+ \sum_{i=1}^I \hat p_i^2 =
\sum_{i=1}^I (\hat  p_i - p_i)^2=
\frac{1}{n^2}\sum_{i=1}^I (n_i-n_i^*)^2.
$$
We will now derive the asymptotic null distribution of $V$-statistic $\mathcal E_n$ from classical $U-$statistic theory (our $V$-statistic is a $U$-statistic plus an asymptotically constant term). By conveniently working out expression~\ref{def:E:app}, we get:
$$
\mathcal E_n/n=\frac{1}{n^2}\sum_{l,m=1}^n\left[ -d(x_l,x_m) +\E d(x_l,X) +\E d(x_m,X)-\E d(X,X') \right] \equiv \frac{1}{n^2}\sum_{l,m=1}^n h(x_l,x_m);
$$
where we define $h$ as the symmetric function:
$$h(y,z):=-d(y,z) +\E d(y,X) +\E d(z,X)-\E d(X,X').$$

By grouping the terms:
$$\mathcal E_n/n=\frac{1}{n^2}\sum_{l\neq m}h(x_l,x_m)+\frac{1}{n^2}\sum_{l=1}^n\E d(x_l,X)-\frac{1}{n}\E d(X,X').
$$
Now multiplying both sides by $n$, the following expression for the energy distance arises:
\begin{equation}\label{E:for:asym}
\mathcal E_n=\frac{n(n-1)}{n^2}\:n\,\mathcal U + \frac{1}{n}\sum_{i=1}^I\hat p_i(1-p_i)-\E d(X,X').
\end{equation}
Applying the unnumbered theorem in Section 5.5.2 of \citet{Serfling}, we see that
$$ n\,\mathcal U\distrilim \sum_{i=1}^I  \lambda_i (Z_i^2-1) $$
as $n\to\infty$, where we note that $\mathcal U=\frac{1}{n(n-1)}\sum_{l\neq m}h(x_l,x_m)$ is a $U$-statistic and $\{\lambda_i\}_i$ is the spectrum of matrix
$$
\bC=(p_i \delta_{ij} - p_i p_j)_{I\times I}.
$$
Summing the elements of its diagonal yields its trace:
$$
\tr(\bC)=\sum_{i=1}^I (p_i-p_i^2)=1-\sum_{i=1}^I p_i^2=\E d(X,X') .
$$
We finally see that the middle term in~(\ref{E:for:asym}) converges in distribution to $0$ under the null, owing to the fact that $\hat p_i\aslim p_i$ by the strong law of large numbers. In conclusion:
$$
\mathcal E_n \distrilim \sum_{i=1}^I  \lambda_i (Z_i^2-1) + \sum_{i=1}^I  \lambda_i=
\sum_{i=1}^I  \lambda_i Z_i^2,
$$
where $\{Z_i^2\}_{i=1}^I$ are IID chi-squared variables with one degree of freedom each.\qed

\begin{thebibliography}{10}
\bibitem[Agresti(2019)]{Agresti}{Agresti, A. G.} (2019). \textit{An Introduction to Categorical Data Analysis}. 3rd edition. John Wiley \& Sons.

\bibitem[Berrett \textit{et al.}(2021)]{BKS}{Berrett, T. B., Kontoyiannis, I. and Samworth, R. J.} (2021). Optimal rates for independence testing via $U$-statistic permutation tests. \textit{Annals of Statistics} \textbf{49,} 2457--2490.

\bibitem[Berrett and Samworth(2021)]{BS}{Berrett, T. B. and Samworth, R. J.} (2021). USP: An independence test that improves on Pearson’s chi-squared and the \textit{G}-test. \textit{Proceedings of the Royal Society (Series A)} \textbf{477,} article 2021.0549.

\bibitem[Castro-Prado \textit{et al.}(2026)]{Epistasis:paper}{Castro-Prado, F., Costas, J., Edelmann, D., Gonz\'alez-Manteiga, W. and Penas, D. R.} (2026). Testing for genetic interaction with distance correlation. [Accepted at the \textit{Biometrical Journal}.] \url{https://arxiv.org/abs/2012.05285v2}.

\bibitem[Castro-Prado and Gonz\'alez-Manteiga(2020)]{FW}{Castro-Prado, F. and Gonz\'alez-Manteiga, W.} (2020). Nonparametric independence tests in metric spaces: What is known and what is not. \url{https://arxiv.org/abs/2009.14150}.

\bibitem[de Wet(1987)]{de Wet}{de Wet, T.} (1987). Degenerate U- and V-statistics. \textit{South African Statistical Journal} \textbf{21,} 99--129.

\bibitem[de Wet(1987)]{Wet:Randles}{de Wet, T. and Randles, R. H.} (1987). On the effect of substituting parameter estimators in limiting $\chi^2$ $U$ and $V$ statistics. \textit{Annals of Statistics} \textbf{15,} 398--412.

\bibitem[Duchesne and Lafaye de Micheaux(2010)]{Duchesne}{Duchesne, P. and Lafaye de Micheaux, P.} (2010). Computing the distribution of quadratic forms: Further comparisons between the Liu-Tang-Zhang approximation and exact methods. \textit{Computational Statistics and Data Analysis} \textbf{54,} 858--862.

\bibitem[Edelmann \textit{et al.}(2024)]{GWAS:paper}{Edelmann, D., Castro-Prado, F. and Goeman, J. J.} (2024). A distance covariance approach to genome-wide association studies. \url{https://arxiv.org/abs/2501.02403}.

\bibitem[Edelmann and Goeman(2022)]{DJ}{Edelmann, D. and Goeman, J. J.} (2022). A regression perspective on generalized distance covariance and the Hilbert--Schmidt independence criterion. \textit{Statistical Science} \textbf{37,} 562--579.

\bibitem[Facal \textit{et al.}(2022)]{Facal:Scand}{Facal, F., Arrojo, M., Paz, E., Páramo, M. and Costas, J.} (2022). Association between psychiatric hospitalizations of patients with schizophrenia and polygenic risk scores based on genes with altered expression by antipsychotics. \textit{Acta Psychiatrica Scandinavica} \textbf{146,} 139--150.

\bibitem[Farebrother(1984)]{Farebrother}{Farebrother, R. W.} (1984). Algorithm AS 204: The distribution of a positive linear combination of chi-squared random variables. \textit{Journal of the Royal Statistical Society: Series C (Applied Statistics)} \textbf{33,} 332--339.

\bibitem[Fisher(1934)]{FET}{Fisher, R. A.} (1934). \textit{Statistical Methods for Research Workers}. 5th edition. Oliver and Boyd.

\bibitem[Goeman \textit{et al.}(2011)]{Jelle:Biomet}{Goeman, J. J., van Houwelingen, H. C. and Finos, L.} (2011). Testing against a high-dimensional alternative in the generalized linear model: Asymptotic type I error control. \textit{Biometrika} \textbf{98,} 381--390.

\bibitem[Hardy(1908)]{Hardy}{Hardy, G. H.} (1908). Mendelian proportions in a mixed population. \textit{Science} \textbf{28,} 49--50.

\bibitem[Imhof(1961)]{Imhof}{Imhof, J. P.} (1961). Computing the distribution of quadratic forms in normal variables. \textit{Biometrika} \textbf{48,} 419--426.

\bibitem[Jakobsen(2017)]{Jakobsen}{Jakobsen, M. E.} (2017). Distance covariance in metric spaces: Non-parametric independence testing in metric spaces. \url{https://arxiv.org/abs/1706.03490v1}.

\bibitem[Jimémez-Gamero \textit{et al.}(2003)]{Gamero}{Jimémez-Gamero, M. D., Muñoz-García, J. and Pino-Mejías, R.} (2003). Bootstrapping parameter estimated degenerate $U$ and $V$ statistics. \textit{Statistics and Probability Letters} \textbf{61,} 61--70.

\bibitem[Lee(1990)]{Lee}{Lee, A. J.} (1990). \textit{$U$-statistics: Theory and Practice}. 1st edition. Springer.

\bibitem[Lyons(2013)]{Lyons}{Lyons, R.} (2013). Distance covariance in metric spaces. \textit{Annals of Probability} \textbf{41,} 3284--3305.

\bibitem[Nassar \textit{et al.}(2023)]{UCSC}{Nassar, L., Barber, G., Benet-Pagès, A., Casper, J., Clawson, H., Diekhans, M.} et al. (2023). The UCSC Genome Browser database: 2023 update. \textit{Nucleic Acids Research} \textbf{51,} 91--97. Online resource available at: \url{https://genome.ucsc.edu/index.html} .

\bibitem[Patefield(1981)]{Patefield}{Patefield, W. M.} (1981). Algorithm AS 159: An efficient method of generating $r \times c$ tables with given row and column totals. \textit{Applied Statistics} \textbf{30,} 91--97. Code available at: \url{https://people.sc.fsu.edu/~jburkardt/m_src/asa159/asa159.html} .

\bibitem[Pearson(1900)]{Pearson}{Pearson, K.} (1900). On the criterion that a given system of deviations from the probable in the case of a correlated system of variables is such that it can be reasonably supposed to have arisen from random sampling. \textit{Philosophical Magazine (Series 5)} \textbf{50,} 157--175.

\bibitem[Preisser and Koch(1997)]{Preisser}{Preisser, J. and Koch, G.} (1997). Categorical data analysis in public health. \textit{Annual Review of Public Health} \textbf{18,} 51--82.

\bibitem[Rizzo and Székely(2016)]{ED}{Rizzo, M. L. and Székely, G. J.} (2016). Energy distance. \textit{Wiley Interdisciplinary Reviews: Computational Statistics} \textbf{8,} 27--38.

\bibitem[Sejdinovic \textit{et al.}(2013)]{Sejdinovic}{Sejdinovic, D., Sriperumbudur, B., Gretton, A. and Fukumizu, K.} (2013). Equivalence of distance-based and RKHS-based statistics in hypothesis testing. \textit{The Annals of Statistics}, \textbf{41,} 2263--2291.

\bibitem[Serfling(1980)]{Serfling}{Serfling, R. J.} (1980). \textit{Approximation Theorems of Mathematical Statistics}. 1st edition. John Wiley \& Sons.

\bibitem[Székely and Rizzo(2005)]{N}{Székely, G. J. and Rizzo, M. L.} (2005). A new test for multivariate normality. \textit{Journal of Multivariate Analysis} \textbf{93,} 58--80.

\bibitem[Székely and Rizzo(2017)]{TEOD}{Székely, G. J. and Rizzo, M. L.} (2017). The energy of data. \textit{\mbox{Annual} Review of Statistics and Its Application} \textbf{4,} 447--479.

\bibitem[Székely \textit{et al.}(2007)]{SRB}{Székely, G. J., Rizzo, M. L. and Bakirov, N.} (2007). Measuring and testing dependence by correlation of distances. \textit{Annals of Statistics} \textbf{35,} 2769--2794.

\bibitem[Torkamani \textit{et al.}(2018)]{PRS:Topol}{Torkamani, A., Wineinger, N. and Topol, E.} (2018). The personal and clinical utility of polygenic risk scores. \textit{Nature Reviews Genetics} \textbf{19,} 581--590.

\bibitem[Trubetskoy \textit{et al.}(2022)]{PGC3}{Trubetskoy, V., Pardiñas, A. F., Qi, T., Panagiotaropoulou, G., Awasthi, S., Bigdeli, T. B.} et al. (2022). Mapping genomic loci implicates genes and synaptic biology in schizophrenia. \textit{Nature} \textbf{604,} 502--508.

\bibitem[Weinberg(1908)]{Weinberg}{Weinberg, W.} (1908). Über den Nachweis der Vererbung beim Menschen. \textit{Jahreshefte des Vereins für vaterländische Naturkunde in Württemberg} \textbf{64,} 368--382.

\end{thebibliography}
\end{document}